# Elucidating Structure Formation in Highly Oriented Triple Cation Perovskite Films

*Oscar Telschow, Niels Scheffczyk, Alexander Hinderhofer, Lena Merten, Ekaterina Kneschaurek, Florian Bertram, Qi Zhou, Markus Löffler, Frank Schreiber, Fabian Paulus and Yana Vaynzof\**

O. Telschow, Q. Zhou, Y. Vaynzof

Integrated Center for Applied Physics and Photonic Materials, Technische Universität Dresden, Nöthnitzer Straße 61, 01187 Dresden, Germany

O. Telschow, Q. Zhou, F. Paulus, Y. Vaynzof

Center for Advancing Electronics Dresden (cfaed), Technische Universität Dresden, Helmholtzstraße 18, 01069 Dresden, Germany

N. Scheffczyk, A. Hinderhofer, L. Merten, E. Kneschaurek, F. Schreiber

Institut für Angewandte Physik, Universität Tübingen, 72076 Tübingen, Germany

F. Bertram

Deutsches Elektronen-Synchrotron DESY, Notkestr. 85, 22607 Hamburg, Germany

M. Löffler

Dresden Center for Nanoanalysis (DCN), Technische Universität Dresden, Helmholtzstraße 18, 01069 Dresden, Germany

Metal halide perovskites are an emerging class of crystalline semiconductors of great interest for application in optoelectronics. Their properties are dictated not only by their composition, but also by their crystalline structure and microstructure. While significant efforts were dedicated to the development of strategies for microstructural control, significantly less is known about the processes that govern the formation of their crystalline structure in thin films, in particular in the context of crystalline orientation. In this work, we investigate the formation of highly oriented triple cation perovskite films fabricated by utilizing a range of alcohols as an antisolvent. Examining the film formation by in-situ grazing-incidence wide-angle X-ray scattering reveals the presence of a short-lived highly oriented crystalline intermediate, which we identify as $FAI-PbI_2-xDMSO$. The intermediate phase templates the crystallisation of the perovskite layer, resulting in highly oriented perovskite layers. The formation of this DMSO containing intermediate is triggered by the selective removal of DMF when alcohols are used as an antisolvent, consequently leading to differing degrees of orientation depending on the antisolvent properties. Finally, we demonstrate that photovoltaic devices fabricated from the highly oriented films, are superior to those with a random polycrystalline structure in terms of both performance and stability.

## 1. Introduction

Metal halide perovskites are a remarkable class of semiconductors whose excellent optoelectronic properties make them particularly promising for application in photovoltaics.[1-2] Over the last decade, significant advances have been made in the design of their composition,[3] passivation of defects,[4] interfacial engineering,[5] and control over the layer microstructure.[6] The latter has been shown to be highly important, since the microstructure of the perovskite active layer has a significant effect not only on the optoelectronic properties and device efficiency,[7] but also on the stability of perovskite solar cells.[8-9]

Many different strategies have been employed for controlling the microstructure of the perovskite layer. For example, increasing the precursor concentration of the perovskite solution has been shown to lead to an increase in the grain size of the perovskite layer.[10-11] Alternatively, Lee *et al.* demonstrated that the microstructure of the perovskite active layer evolves when chlorine-containing precursors such as lead chloride ($PbCl_2$) or methylammonium chloride (MACl) are added to the perovskite solution, leading to significantly larger grain sizes.[12] Moreover, the use of additives also proved effective in controlling the microstructure of perovskite layers. Notable examples of such additives are thiourea,[13] ammonium hypophosphite ($NH_4H_2PO_2$)[14] and hypophosphorous acid.[15] Finally, the surface properties of the substrate on top of which the perovskite precursor solution is deposited can also impact on the resultant microstructure. Non-wetting surfaces have been shown to lead to the formation of larger grains,[16] but on the other hand might also result in microstructural defects such as pin-holes and nanovoids.[17]

In addition to the size of the perovskite grains, recent studies suggest that their relative orientation with respect to the substrate and each other might impact on the photovoltaic performance.[18] For example, Yang *et al.* reported that the addition of caffeine into the perovskite solution results in a preferential orientation of the perovskite grains along the (110) planes. The authors suggest that this preferential orientation improves charge transport in the device, leading to enhanced photovoltaic performance.[19] However, the impact of grain orientation could not be disentangled in this case, since the addition of caffeine also led to increase in grain size and defect passivation, which also result in improved device performance.

Interestingly, the work examining the impact of chlorine also reported a preferential orientation along the (110) upon the addition of MACl into the perovskite precursor solution.[12] Yet, also in this case, the impact of change of orientation could not be decoupled from the change in microstructure. On the other hand, spectroscopic studies suggest that neither the size, nor the orientation of perovskite grains impacts their optoelectronic properties,[20] leaving the question of the consequences of crystalline orientation for device performance unanswered.

Importantly, the formation of perovskite films occurs via crystalline intermediate phases, often containing high boiling point solvent molecules in the crystal lattice.[21] These intermediate phases convert to the photoactive perovskite phase upon thermal annealing.[22-23] Understanding the formation mechanisms of such intermediates, and the development of strategies to control them can enable precise structural engineering of the deposited perovskite layers.[24] Among the most effective methods to investigate the temporal evolution of the crystallization process is by in-situ grazing-incidence wide-angle X-ray scattering (GIWAXS) measurements.[25-27] Indeed, such characterization – although experimentally complex – has already led to significant insights. For example, Qin *et al.* identified three clear stages of film formation of mixed perovskites, and demonstrated that annealing has to take place in the second stage, in order to avoid the formation of undesirable phases.[28] Huang and co-workers employed in situ X-ray diffraction (XRD) to investigate the crystallization processes in $FAPbI_3$ and demonstrated the presence of multiple solvent-coordinated intermediate phases.[29] While these examples illustrate the efficacy of in-situ characterization for the study of perovskite crystallization processes, to the best of our knowledge, these techniques were not yet applied to the study of orientation control.

To examine the impact of orientation on the photovoltaic performance, it is thus important to not only isolate the orientational variation from microstructural changes, but also investigate the temporal evolution of structure formation, thus elucidating the mechanism that triggers orientational preference. In our previous work, we observed that the former can be made possible in case of perovskite layers fabricated via the antisolvent engineering route.[30] Specifically, we observed that the use of alcohols as antisolvent leads to highly oriented films, while other antisolvents largely lead to a random grain orientation. A similar observation was later reported by Wang *et al*, who observed preferred orientation of perovskite layers fabricated using isobutanol (IBA) as an antisolvent.[31] The authors suggested that the polarity of the antisolvent molecule led to a different orientation of formamidinium ($FA^+$) molecules in an $IBA$-$DMSO$-$FA^+$ complex as compared to the $DMSO$-$FA^+$ complexes formed when using a non-alcoholic antisolvent. Importantly, the authors observed an improved photovoltaic performance for the oriented perovskite layers. While these results are highly promising, many questions regarding the structure formation of oriented perovskite films and the impact on the photovoltaic performance remain open. For example, it remains unclear which characteristics of the alcoholic antisolvents impact the orientation of the perovskite layers and how the relative degrees of orientation impact the performance and stability of perovskite solar cells.

To address these questions, we investigate the temporal evolution of crystallization in triple cation perovskite films deposited by antisolvent engineering method. In short, in this method the perovskite thin film is formed by spin-coating the perovskite solution (in a 4:1 mixture of *N,N*-dimethylformamide (DMF) and dimethyl sulfoxide (DMSO)) on to the substrate, during which an antisolvent is dripped onto the substrate, triggering crystallization. Once the spin-coating procedure ends, the crystallization is completed by thermal annealing (**Figure 1a**). To probe the structure formation, we employed in-situ grazing-incidence wide-angle X-ray scattering (GIWAXS) performed during the fabrication of the perovskite layers. Such in-situ techniques proved to be highly effective in studying the crystallization processes of perovskite films, revealing both crystallization kinetics and growth mechanisms.[32] We examine the structure formation of perovskite films fabricated using three different alcoholic antisolvents comparison to films fabricated using a non-alcoholic solvent. The chemical structures of the examined antisolvents, namely butanol (BuOH), isopropanol (IPA), isobutanol (IBA) and trifluorotoluene (TFT) are shown in **Figure 1b**. Our measurements reveal the presence of a short-lived, highly oriented intermediate species that templates the growth of the oriented perovskite layer. Finally, we compare the performance and stability of the fabricated perovskite solar cells, revealing that both these factors are correlated with the degree of crystal grain orientation.

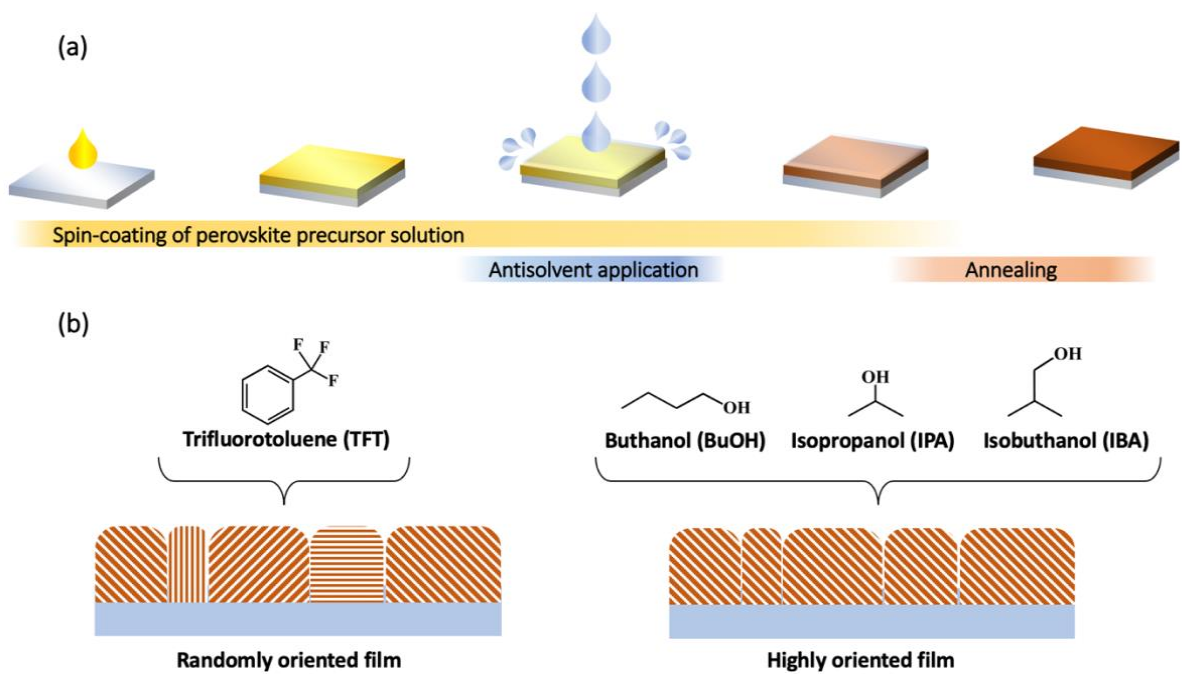

**Figure 1.** (a) Schematic illustration of the antisolvent engineering method for perovskite layer deposition. (b) Chemical structures and illustration of the corresponding film orientation of the antisolvents used in this study.

## 2. Results and Discussion

### 2.1. Microstructure Characterization

In our previous work we reported that the use of alcohols as antisolvents may lead to microstructural defects due to the extraction of organic halides during the antisolvent application step.[30] Specifically, this may occur if the antisolvent is extruded slowly and the interaction time of the antisolvent with the spinning substrate and thinned precursor solution is not short enough. To avoid this, all films in this study were fabricated by extruding the antisolvents rapidly. To ensure no microstructural defects were formed, the films were examined using scanning electron microscopy (SEM). SEM images confirmed that polycrystalline and pinhole-free perovskite films were fabricated using each of the antisolvents (**Figure 2**). This has been further corroborated via cross-sectional SEM imaging that confirmed that all antisolvents led to the formation of compact perovskite layers without any pin-holes or nanovoids at the buried interfaces (**Supplementary Information Figure S1**). The grain size is similar in all the films with grains ranging from 50 to 300 nm in diameter. Interestingly, the change in the relative orientation in these films can be observed already via SEM. While grains in perovskite films fabricated using TFT exhibit various edges of crystal facets, grains on the IPA, IBA and BuOH films often display a flat surface facing upward with concentric edges around them, suggesting crystal planes parallel to the film surface. Some of these edges have been highlighted in the high magnification images shown in Figure 2 for clarity.

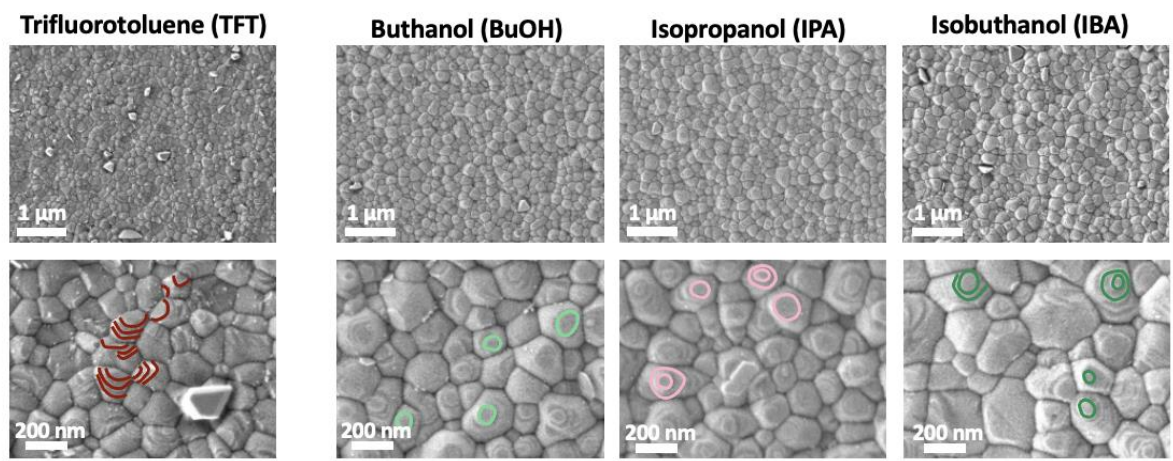

**Figure 2.** Scanning electron microscopy images collected via the secondary electron detector of triple cation perovskite films fabricated using different antisolvents. Exemplary grain edges have been highlighted to show the difference in orientation.

**2.2. Structural Characterization**

To examine in detail the evolution of the crystalline structure during the formation of the films, in-situ GIWAXS characterisation was performed on a bespoke setup in which a spin-coater was integrated into the synchrotron beamline and the reciprocal space maps were recorded as a function of time for each of the investigated antisolvents. The evolution of these maps as videos can be found as **Supplementary Note 1**.

In **Figure 3** we present the GIWAXS maps at important time points during the deposition procedure for samples with IPA as an antisolvent, which will allow us to track the evolution of different structures formed during the film formation. At the first frame (15s), taken after the perovskite precursor solution was dispensed, but prior to the start of spin-coating, no crystalline features are observed. The spin-coating procedure was started at 37s, and shortly afterwards (50s), the solution is thinned down, making it possible to observe the reflections associated with the ITO substrate, which is marked in a dashed line (q = 2.15 Å$^{-1}$). We note that the feature at 0.49 Å$^{-1}$ originates from the kapton window of the experimental setup and is present on all images independent from the sample properties. The antisolvent is dispensed at 67s, at which point immediately two crystalline species can be observed. The first, marked in yellow circles, leads to a strong signal at $q_z$ = 0.54 Å$^{-1}$, which we assign to a highly oriented intermediate phase that templates the oriented growth of the perovskite, since - as will be shown in the following - it is only observed in the case of the alcoholic antisolvents. The intermediate species is very short lived and is observed only for 2 seconds under the applied preparation methods. The second species, marked in orange circles is a hexagonal phase of the triple cation perovskite.[33] It is noteworthy that already at this stage the hexagonal phase exhibits a clearly preferred orientation, since distinct diffraction features are observed, rather than full diffraction rings. Shortly after the spin-coating has finished (100s), we observe a co-existence of the hexagonal and cubic phases (marked in black circles) of the perovskite layer.[34] The latter also exhibits a highly oriented structure, evidenced by distinct diffraction features. Due to instrumental limitations, annealing could only commence roughly 2 minutes after the completion of spin-coating. Approximately 100s after spin coating stopped (199s), we no longer observe a hexagonal phase of the perovskite, but instead detect the formation of a known $(MA)_2Pb_3I_8 \cdot 2DMSO$ intermediate (pink circles).[35] This intermediate remains for the first 20s of annealing, but is eliminated after 90s of annealing, at which point a small contribution associated with phase separated $PbI_2$ can be observed (red circle) alongside highly oriented features of cubic perovskite.

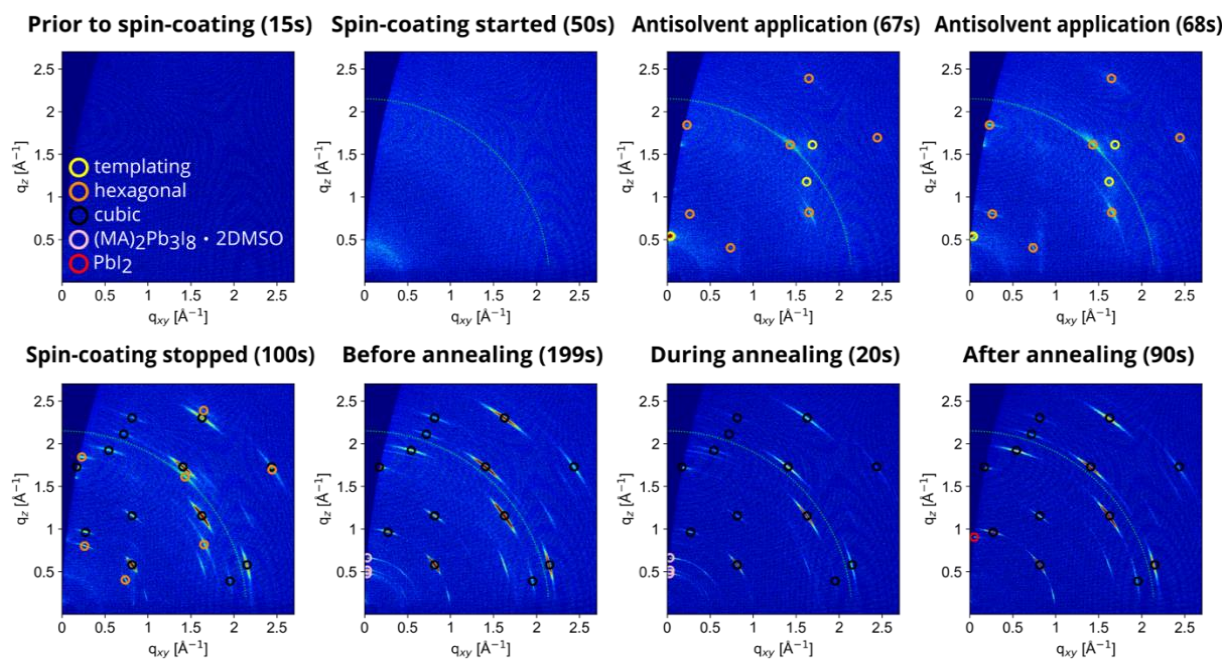

**Figure 3**: GIWAXS data taken during different time points during film formation using IPA as an antisolvent. The evolution for the other antisolvents can be seen in the Supplementary Information as **Figures S2-S4**.

To compare the structure evolution for the different antisolvents, we focus our attention on the templating species and hexagonal and cubic phases of the perovskite. **Figure 4** displays the intensity evolution of these three species as a function of time and the final GIWAXS maps obtained post annealing for each of the films. In case of the TFT antisolvent, no templating species is observed and the formation of the hexagonal perovskite phase – which, in contrast to the alcohols, shows significantly less orientation - occurs once the antisolvent is dispensed. After an initial increase, this phase is decreased with an increasing intensity of the cubic phase. Once annealed, only cubic phase features remain, with a largely random orientation, evidenced by the Debye ring shape of the GIWAXS pattern. On the other hand, in the case of all three of the alcoholic antisolvents, a short-lived templating species is observed immediately once the antisolvent is dispensed, which in all cases appeared at $q_z = 0.54$ Å$^{-1}$. This finding suggests that the structure of this species is independent of the specific alcohol used, which implies that the antisolvent is not incorporated into that crystalline structure.

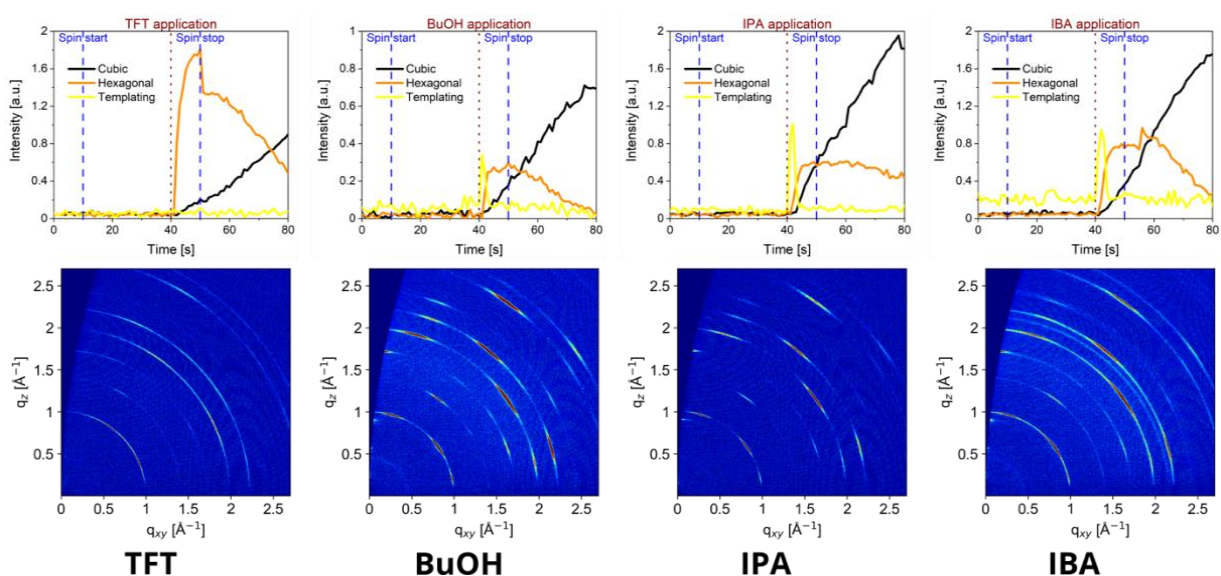

**Figure 4**: Temporal evolution of the intensity of the templating species, the hexagonal and cubic perovskite phases and the final GIWAXS map obtained post annealing for each the investigated antisolvents by integrating the relevant diffraction rings for each of the tracked species.

To compare the degrees of orientation between the different samples, we performed angular integration along the (100) reflection as well as X-ray diffraction (XRD) measurements that enable us to compare the intensity of the (111) reflections (q = 1.72 Å$^{-1}$, 2Θ = 24.46°). The angular profiles confirm that no preferential orientation in the case of the TFT samples, but a clearly preferred orientation for all the alcoholic antisolvents with increased intensity at approximately 15°, 55° and 77° (**Figure 5a**). The distribution of intensities shows a dependency on the choice of antisolvent, with IPA leading to particularly oriented films with the strongest intensity at 55° in comparison to that at 15° and 77°. Similar observations can be made by examining the XRD patterns (**Figure 5b**). Very clearly, the (111) intensity is strongest in the IPA fabricated samples, although it is very prominent also in the other samples fabricated with alcoholic antisolvents.

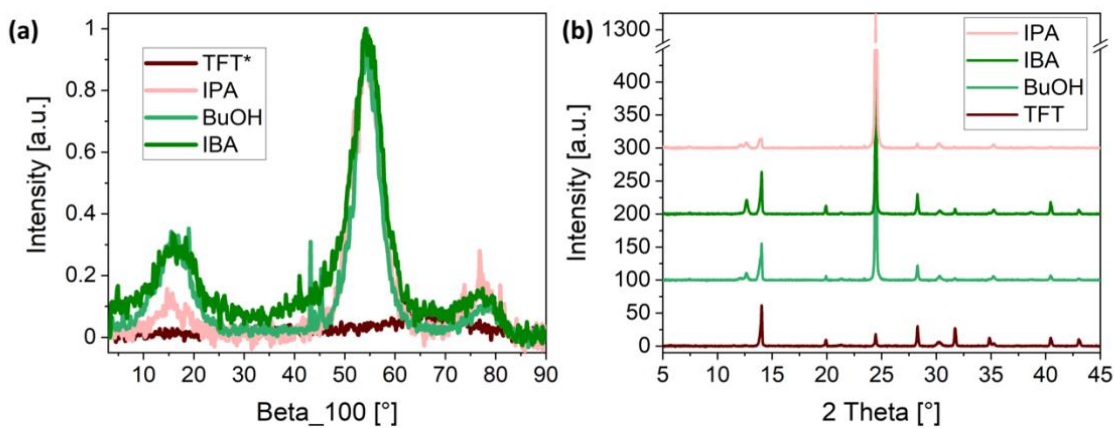

**Figure 5**. (a) Angular profile along the (100) reflection of the GIWAXS maps shown in Figure 4. (b) XRD measurements on perovskite samples fabricated using different antisolvents. * For clarity of the graph, the TFT curve was normalized to the highest peak of BuOH instead of its own highest peak.

To investigate whether there are differences in the vertical distribution of the crystalline species in the fully fabricated, annealed perovskite layers, we performed angular dependent GIWAXS measurements. At this stage, the films consist almost exclusively of the cubic perovskite phase and $PbI_2$, the vertical evolution of which is shown in **Supplementary Information Figure S5**. The results reveal that the vertical distribution of the $PbI_2$ is impacted by the choice of antisolvent: while alcoholic antisolvents lead to increased amounts of $PbI_2$ in the bulk of the films. In the films made using TFT, its distribution is homogenous throughout the layers. In the past, we and others reported that eliminating $PbI_2$ from the sample surface of triple cation perovskites leads to improved performance, in particular due to an increase in the device open-circuit voltage ($V_{OC}$).[36-38] The potential impact of the differences in the $PbI_2$ distribution on the device performance will be discussed in Section 2.4.

## 2.3. Proposed Mechanism for Structure Formation

As mentioned above, our previous observation that the short-lived templating structure appears at $q_z = 0.54$ Å$^{-1}$ regardless of the type of alcohol used suggests that the alcoholic antisolvent is not integrated into this crystalline structure, indicating that it consists of the precursors and/or solvents present in the wet perovskite film being spin-coated. Its reflections do not coincide with the previously reported solvent complexes involving DMF or $PbI_2$-DMSO.[35-41] To gain further insights into the species that are integrated into the short-lived intermediate phase, we examined the structure formation in MAPbI$_3$ and FAPbI$_3$ films fabricated with IPA as an antisolvent. Interestingly, GIWAXS measurements revealed that only the latter composition exhibited a templating structure (**Supplementary Figure S6**). This observation suggests that FA molecules are incorporated into the templating structure, since the high degree of orientation depends significantly on the FA content. Drop-casting a highly concentrated solution of FAI and PbI$_2$ in a molar 1:1 ratio in pure anhydrous DMSO resulted after gentle drying at 60°C in a pale yellow, crystalline film that exhibits a series of intense reflections that are in a good agreement with those observed by GIWAXS for the templating structure (see **Supplementary Figure S7**). The crystalline film is highly ordered, which can also be observed via optical microscopy (**Supplementary Figure S8**). We note that single crystal structure characterisation and in-plane diffraction experiments failed due to the high sensitivity of these crystals, that converted to brown perovskites rather rapidly under light or X-ray exposure, which is typical for perovskite intermediates that incorporate solvent molecules. Our experiments suggest a composition of FAI-PbI$_2$-x·DMSO considering the 1:1 FAI to PbI$_2$ ratio we used. Examining the literature reveals that an intermediate with this composition has been proposed by Ren *et al*,[42] yet its complete crystalline structure has not been reported by the authors. The absence of DMF in the drop-casting experiments proves that solely DMSO is incorporated into the crystal lattice of the observed intermediate upon treatment with alcoholic antisolvents. This suggests that alcoholic antisolvents are preferentially removing DMF from the DMF:DMSO host solvent mixtures used in the film fabrication. This hypothesis is supported by considering the Hansen solubility parameters of the solvents involved in the film fabrication process. Hansen solubility parameters, established by Charles M. Hansen in 1967,[43] are defined as follows:

$\Delta$D - The energy from dispersion forces between molecules

$\Delta$P - The energy from dipolar intermolecular force between molecules

$\Delta$H - The energy from hydrogen bonds between molecules.

Hansen defined the "Hansen space" as the three-dimensional coordinate space ($\Delta$D, $\Delta$P, $\Delta$H). The closer two molecules are to each other in this Hansen space, the more likely it is that they are capable of dissolving in each other. **Table 1** lists the Hansen solubility parameters for the host solvents (DMF and DMSO) and the four antisolvents used in this study. Based on these parameters, we can calculate the distance between the corresponding coordinates in the Hansen space for each of the antisolvents with respect to the host solvents, defined as R$_A$(DMF) and R$_A$(DMSO). By examining these distances, we observe that the interaction of DMF with the alcoholic antisolvents is far stronger to that of DMSO, evidenced by the smaller values of R$_A$(DMF). This is due to the stronger interaction via hydrogen bonds that can form between the alcohols and the polar DMF host solvent.

On the other hand, due to the absence of a hydroxyl group in TFT, only the dispersive and dipolar interactions determine the interaction among the solvents. These interactions are very similar for both host solvents, resulting in an equally good extraction of both DMF and DMSO by TFT. This is further illustrated by calculating the R$_A$(DMF)/R$_A$(DMSO) ratio, which is significantly smaller for the alcohols than for TFT. This difference in solvent interaction suggests that alcohols preferentially extract DMF from the precursor solution as it is more soluble in them, resulting in a local enrichment of DMSO on the substrate during the antisolvent treatment. The high DMSO concentration, in turn, enables the formation of the highly oriented FAI-PbI$_2$-xDMSO intermediate that templates the crystallisation of the perovskite and consequently its orientation. Employing TFT – or other non-alcoholic solvents - as antisolvent does not lead to a DMSO enriched environment and the DMSO-intermediate does not form, thus leading to a lack of preferred orientation in the final perovskite film.

**Table 1**: Hansen parameters of the perovskite solvents and antisolvents used in this study.

| Solvent | Δ D | Δ P | Δ H | R$_A$(DMF) | R$_A$(DMSO) | R$_A$(DMF)/R$_A$(DMSO) |
|---|---|---|---|---|---|---|
| IPA | 15.8 | 6.1 | 16.4 | 9.7 | 13.10 | 0.74 |
| BuOH | 16 | 5.7 | 15.8 | 9.6 | 13 | 0.74 |
| IBA | 15.1 | 5.7 | 15.9 | 10.31 | 13.31 | 0.75 |
| TFT | 17.5 | 8.8 | 0 | 12.32 | 12.85 | 0.96 |
| DMF | 17.4 | 13.7 | 11.3 | 0 | 3.54 | - |
| DMSO | 18.4 | 16.4 | 10.2 | 3.54 | 0 | - |

Taken together with the results of our previous studies,[30,44] the proposed mechanism adds an additional consideration into the selection of an antisolvent for perovskite film fabrication, resulting in three different factors that impact film formation:

(1) The solubility of the perovskite precursors in the antisolvent: in case the chosen antisolvent can easily dissolve some of the perovskite precursors, its application may lead to an irreparable alternation of the intended film stoichiometry. This can be largely avoided by applying the antisolvent very fast,[30] or modifying its deposition strategy from pipetting to spraying.[38]

(2) The miscibility of the antisolvent with the host solvents: certain antisolvents exhibit a very poor miscibility with the host solvents DMF and DMSO. In this case, the extraction of the host solvents is inefficient, often resulting in an incomplete film coverage.[30,44] To circumvent this issue, the antisolvent should be applied slowly in order to prolong the time during which the host antisolvents can be extracted.

(3) The solubility of the antisolvent with the host solvents: considering that many perovskite compositions rely on the use of host solvent mixture (e.g. DMF and DMSO), the individual interactions of the antisolvent with each of the host solvents must be considered. Differences in the solubility of the antisolvent with each of the host solvents may lead to a preferential extraction of one of the over the other, which in turn can impact on the formation of solvent-containing intermediate phases that guide perovskite crystallisation.

These three factors have to be considered together when selecting the antisolvent, and are influenced by the specific perovskite composition, the desired stoichiometry, microstructure and orientation. Moreover, the considerations outlined above can be used to choose a mixture of antisolvents that would lead to the desired film formation processes. For example, we fabricated perovskite layers using a 1:1 mixture of TFT and IPA. Such a sample exhibits preferential orientation as the one fabricated by IPA (**Supplementary Figure S9**), but the presence of TFT lowers the solubility of the perovskite precursors in the antisolvent mixture, thus relaxing the need to apply it very fast. Importantly, the use of antisolvent mixtures also opens the possibility to utilize them as a mean to incorporate additives or passivation agents to the perovskite surface.[17,45]

### 2.4. Photovoltaic Characterization

To investigate the performance of perovskite layers fabricated with the different antisolvents, we fabricated solar cells in an inverted architecture, with the structure glass/ITO/MeO-2PACz /perovskite/PCBM/BCP/Ag. The photovoltaic parameters of the best 6 solar cells of each kind are presented in **Figure 6**. The $V_{OC}$ and fill factor (FF) of the devices are very similar, but the short-circuit current ($J_{SC}$) shows clear difference, with devices fabricated with TFT as antisolvent yielding the lowest average photocurrent. This observation is in agreement with external quantum efficiency (EQE) measurements that show a higher yield for devices made by alcoholic antisolvents as compared to that of TFT (**Supplementary Figure S10**). The resulting power conversion efficiencies (PCEs) are highest for IPA and BuOH, averaging just below 20 %. Thereby, they exceed the average PCEs of TFT and IBA, which are below 19%. Exemplary current-density-voltage (J-V) curves are shown in **Supplementary Figure S11.**

The very similar $V_{OC}$ of the devices suggests that the changes in the vertical distribution of the $PbI_2$ observed in the angular dependent GIWAXS measurements are not the cause of the improved photovoltaic performance. At the same time, it is interesting that the highest performance average was achieved for samples fabricated using IPA, which also exhibited the highest degree of preferred orientation. Considering that the improvement is associated with an enhancement in the photocurrent and not in the other solar cell parameters, it is unlikely that it originates from a change in the optoelectronic properties of the layers. Indeed, UV-vis absorption and photoluminescence measurements (**Supplementary Figure S12**) are similar between all the measured samples. This suggests that the change in orientation mainly impacts the charge transport properties, which appears to be enhanced in the case of the highly oriented films.

To gain initial insights into the degradation behaviour of the different devices, their performance was remeasured 23 days later after being stored unencapsulated in the dark in ambient air. The results are presented in **Supplementary Figure S13**. We observe that the degradation in performance is more severe for TFT-based devices, in comparison to that of those made using alcoholic antisolvents. This is an initial indication that the latter exhibit a slower degradation process, and considering that all other parameters in the device fabrication were kept identical, we preliminarily associate this suppression of degradation with the higher degree of orientation in the perovskite active layers. Monitoring the performance evolution of the devices under continuous illumination (**Supplementary Figure S14a**) reveals that TFT based devices exhibit a significantly stronger burn-in (more than 15% of initial performance) than devices made using alcoholic antisolvents (approximately 5%). This observation is in agreement with recent reports that suggest that highly oriented films result in a superior stability under operational conditions.[46-48] When exposed to thermal stress, however, the devices exhibited identical degradation dynamics (**Supplementary Figure S14b**). These results suggest that orientation plays a significant role in determining the degradation dynamics of perovskite solar cells, nevertheless, a comprehensive study of the impact of orientation on degradation mechanisms of perovskite films is a topic of future investigation and is beyond the scope of the current work.

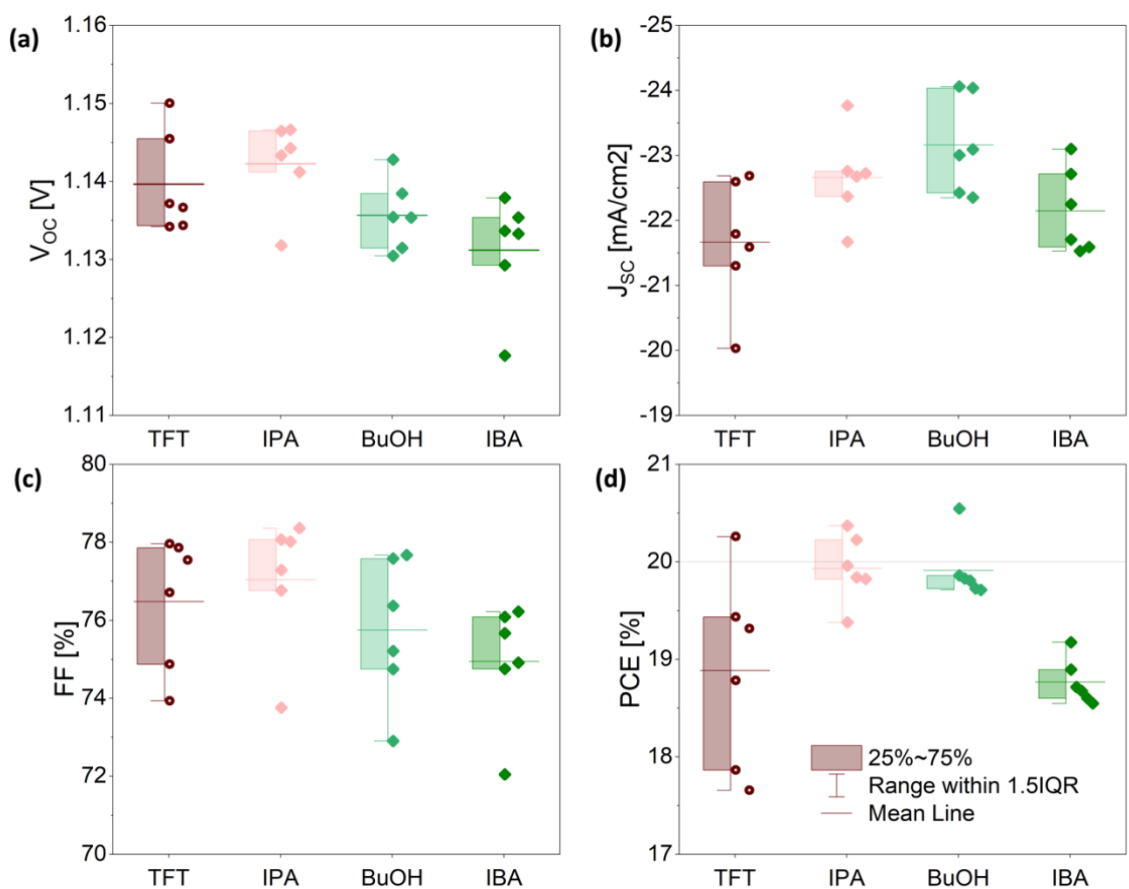

**Figure 6.** (a) $V_{OC}$, (b) $J_{SC}$, (c) FF, (d) PCE of photovoltaic devices fabricated using different antisolvents.

## 3. Conclusion

To summarize, we investigated the film formation processes that govern the growth of highly oriented triple cation perovskite films fabricated by alcoholic antisolvents. By monitoring these processes by in-situ GIWAXS, we uncovered the presence of a highly oriented intermediate species that templates the growth of the perovskite layers. We identify this species to be a FAI-PbI$_{2-x}$·DMSO complex that is formed due to a strong interaction of the alcoholic antisolvents with the DMF host solvent of the perovskites solution, which results in its preferential extraction during the antisolvent application step. We find that films with stronger degree of orientation result in higher photovoltaic performance and stability when incorporated in solar cells, highlighting the importance of developing strategies to control the orientation of polycrystalline perovskite thin films.

## 4. Experimental Section/Methods

*Materials*: Pre-cut glass 12×12 mm$^2$ substrates with a pre-coated central stripe of indium tin oxide (ITO) by Psiotec Ltd. were used as a substrate for device fabrication. Perovskite precursor solution was created with PbI$_2$ and PbBr$_2$ from TCI, CsI from abcr and MAI (CH$_3$NH$_3$I) and FAI (HC(NH$_2$)$_2$I) from GreatcellSolar Materials. PCBM was purchased from Lumtec and MeO-2PACz from TCI. IBA was purchased from Alfa Aesar, EtOH from ACROS Organics and BCP, TFT and all solvents from Sigma Aldrich. The materials, solvents and solutions were stored in a dry nitrogen atmosphere except for PCBM and BCP, which were stored in ambient air. Silver pellets for thermal evaporation of the top contact were purchased from Kurt J. Lesker Company.

*Solution preparation*: MeO-2PACz was used to form a hole-transport layer (HTL). It was dissolved in anhydrous EtOH and the solution was sonicated for 15 min at 30 °C to 40 °C. The 1 mmol L$^{-1}$ solution for spin coating was diluted from a 10 mmol L$^{-1}$ stock solution. The perovskite precursor solutions were prepared in a sequential solution method to keep a precise stoichiometry at 1.2 mol L$^{-1}$ of precursors for Cs$_{0.05}$(MA$_{0.17}$FA$_{0.83}$)$_{0.95}$Pb(I$_{0.9}$Br$_{0.1}$)$_3$ in a 4 : 1 mixture of DMF and DMSO by volume with 1 % excess of PbI$_2$ and 0.25 % ionic liquid ([BMP]$^+$[BF4]$^-$) as additive. In the first step, the component salts were weighed into adequate vials. Then the inorganic salts, CsI, PbI$_2$ and PbBr$_2$, were dissolved in dimethylsulfoxide (DMSO) in the first case and a 4:1 mixture by volume of anhydrous N,N-dimethylformamide (DMF) to DMSO in the two latter cases at 180°C. After the salts had dissolved completely and the solutions had cooled down, the CsI and PbBr$_2$ solutions were added to the PbI$_2$ solution in a volume ratio of 0.05:0.15:0.85 to obtain a 1.2 mol L$^{-1}$ inorganic stock solution of Cs$_{0.05}$PbI$_{1.75}$Br$_{0.3}$ with 1% excess of PbI$_2$. In a molar ratio of 0.95:1 the inorganic stock solution was added into vials with correctly weighed amounts of FAI and MAI. Then, the solution from the MAI vial was added into the FAI solution in a volume ratio of 1:5 MAI to FAI, yielding a 1.2 mol L$^{-1}$ solution of Cs$_{0.05}$(MA$_{0.17}$FA$_{0.83}$)$_{0.95}$Pb(I$_{0.9}$Br$_{0.1}$)$_3$ with 1% excess of PbI$_2$. Finally, the appropriate amount of this solution is transferred to a vial with the ionic liquid [BMP]$^+$[BF4]$^-$, to yield a 0.25% concentration of the organic liquid in the resulting solution. For the electron transport layer (ETL), PCBM was dissolved in anhydrous chlorobenzene (CB) in an amber vial with a concentration of 20 mg mL$^{-1}$. To ensure the dissolution, the mixture was stirred in a nitrogen filled glovebox overnight with a magnetic stirring bar at 70 °C. Afterwards, the solution was filtered through a 0.45 μm PTFE syringe filter. As a hole-blocking layer (HBL), Bathocuproine (BCP) was deposited by means of a 0.5 mg mL$^{-1}$ solution in anhydrous IPA. The solution was prepared by stirring overnight at 70 °C via a magnetic stirring bar under inert atmosphere.

*Device fabrication*: The devices were fabricated in an inverted architecture. Substrates were cleaned by rinsing with acetone and subsequent rinsing with and 7 min sonication at 40°C in soap water, deionized water, acetone and isopropanol. Afterwards, the substrates were blown dry with nitrogen and exposed to an oxygen plasma for 10 min. The HTL and perovskite layer were applied in a humidity-controlled glovebox (GB). To form the HTL as a self-assembled monolayer (SAM), 35 µL of MeO-2PACz solution was spin-coated statically onto the substrate at 3000 RPM for 15 s. The samples were then annealed for 10 min at 100 °C. Perovskite films were fabricated by applying 40 µL of precursor solution before running a two-step spinning program. The sample was first spun at 1000 RPM for 12 s and at 5000 RPM for 28 s afterwards. Antisolvents were applied dynamically 5 s prior to the end of the fast-spinning step in a fast manner and an amount of 150 µL. After the spinning process, the samples were annealed for 30 min at 100 °C. In the In-situ GIWAXS measurements, the antisolvents were applied 10s prior to the end of the fast-spinning step. The ETL and HBL were fabricated in a nitrogen-atmosphere GB. 20 µL of PCBM solution was applied dynamically after 5 s of a 30 s rotation at 2000 RPM. The samples were subsequently annealed for 10 min at 100 °C. After cooling down, the samples were dynamically spin-coated with 40 µL of BCP solution, applied 5 s into a 30 s spinning step at 4000 RPM. An 80 nm thick layer of 99.99 % pure silver (Ag) was thermally evaporated onto the sample locally, to form top contact for the devices. To prevent harm to HBL and ETL, the deposition rate was initially set to 0.01 nm s$^{-1}$ and increased to 0.1 nm s$^{-1}$.

*Photovoltaic characterization*: For the PV performance measurements, an ABET TECHNOLOGIES Sun 3000 AAA solar simulator was used to illuminate the devices with simulated AM 1.5 light under ambient conditions. Currents were measured with a Keithley 2450 SMU. A NIST traceable Si reference cell was used for intensity calibration and corrected by determining the spectral mismatch between solar spectrum, reference cell, and spectral response of the device. Devices contained 8 pixels with an active area of 1.5 mm x 3 mm, which were scanned with a voltage sweep from 1.2 V to 0 V and back with a step size of 0.025V and a dwell time of 0.1 s after 2 s of light soaking at 1.2 V.

*X-Ray diffraction*: The X-ray diffraction measurements were performed on samples containing the device structure up to the perovskite layer in ambient air. The utilized measurement system is a Bruker D8-discover with a Lynxeye 1D detector.

*In-situ GIWAXS characterization*: The in-situ GIWAXS measurements were performed at beamline P08 at PETRA III (DESY, Hamburg).[49] with a photon energy of E = 18 keV and a Perkin Elmer XRD 1621 flat panel detector at a distance of 750 mm. The angle of incidence during in-situ characterization was 0.5° to probe the bulk features of the thin films. To control the application of antisolvents during the experiments, a remote-controlled dispensing system with an attached syringe pump was built into the measurement chamber. Both the spin-coater and the syringe pump are integrated as devices in the beamline control software which allows for electronic synchronization of the spin-coating procedure, antisolvent dispensing and GIWAXS measurement within an error of approximately 1 s. Diffraction intensities were calculated by integrating peak intensities over the entire peak area and applying a baseline correction. This radial integration was performed using the software ImageJ.

*Scanning electron microscopy*: SEM measurements were performed on perovskite films on glass/ITO/MeO-2PACz in vacuum. The pre-patterned ITO stripe was used to ground the samples with silver paste to avoid sample charging. In a ZEISS GeminiSEM 500 the InLens and HE-SE2 detectors were utilized to yield images with a 5 nm resolution with electrons of 1.5 kV landing energy.

*UV-vis absorption*: A Jasco V-770 Spectrophotometer was used to determine the spectral absorption of the perovskite films. The samples were glass substrates coated with perovskite and a pure glass substrate was used as reference. The spectrum was measured form 850 nm to 600 nm with a step size of 1 nm.

*Photoluminescence and PLQE*: To measure PL and PLQE, the samples were fixed in the beam path of a 532 nm laser operated at 5 mW inside a calibrated Labsphere 6 inch QE sphere integration sphere and measured utilizing an Ocean Optics QE65 Pro spectrometer, following the procedure described by De Mello et al.[50] During the measurement, the integration sphere was flushed with nitrogen in order to prevent oxygen or water molecules in ambient air from interacting with the perovskite surface. The samples were glass substrates coated with perovskite.

**Supporting Information**

Supporting Information is available from the Wiley Online Library or from the author.

**Acknowledgements**

This project has received funding from the European Research Council (ERC) under the European Union's Horizon 2020 research and innovation programme (ERC Grant Agreement n° 714067, ENERGYMAPS) and the Deutsche Forschungsgemeinschaft (DFG) in the framework of the Special Priority Program (SPP 2196) project PERFECT PVs (#424216076). We also thank the BMBF for funding (project 05K19V TA).

We acknowledge DESY (Hamburg, Germany), a member of the Helmholtz Association HGF, for the provision of experimental facilities. Parts of this research were carried out at PETRA III at beamline P08. Beamtime was allocated for proposal I-20210858.

Finally, we kindly acknowledge the use of the facilities in the Dresden Center for Nanoanalysis in the cfaed for providing access to the electron microscopy facilities. O. Telschow and N. Scheffczyk contributed equally to this work.